# A Structured Hardware Software Architecture for Peptide Based Diagnosis of Baylisascaris Procyonis Infection

S. M. Vidanagamachchi, S.D. Dewasurendra, R. G. Ragel
Department of Computer Engineering,
University of Peradeniya,
Sri Lanka.

M. Niranjan
Electronics and Computer Science,
University of Southampton,
United Kingdom.

*Abstract*—The problem of inferring proteins from complex peptide cocktails (digestion products of biological samples) in shotgun proteomic workflow sets extreme demands on computational resources in respect of the required very high processing throughputs, rapid processing rates and reliability of results. This is exacerbated by the fact that, in general, a given protein cannot be defined by a fixed sequence of amino acids due to the existence of splice variants and isoforms of that protein. Therefore, the problem of protein inference could be considered as one of identifying sequences of amino acids with some limited tolerance. In the current paper a model-based hardware acceleration of a structured and practical inference approach is developed and validated on a mass spectrometry experiment of realistic size. We have achieved 10 times maximum speed-up in the co-designed workflow compared to a similar software-only workflow run on the processor used for co-design.

*Keywords—proteomics; co-design; Aho-Corasick*

## I. INTRODUCTION

Reliable and rapid inference of proteins from complex samples (containing thousands of different proteins) is a challenging computational problem in shotgun proteomics [1, 2], particularly when the required processing throughput is very high. In protein inference, we have the following challenging and critical data processing tasks: (1) assigning the mass spectra to known peptides and (2) mapping peptides to parent proteins and defining the confidence level of identified proteins [1]. In the present paper we present our contribution to task (2). The solution was formulated around the problem of identifying raccoon roundworm infection starting from a cocktail of peptides resulting from the digestion of biological extracts from affected species. Our protein inference algorithm uses bit-split Aho-Corasick machines adapted for the occurrence of splice variants and isoforms, by considering the problem of protein inference as one of identifying sequences of amino acids with some limited tolerance. In this paper, we present details of a model-based hardware/software co-design methodology for hardware acceleration of the inference work-flow and validate it on an experiment of realistic size. In our previous work [28] we presented the steps of a bottom-up approach for the protein inference process, but it was neither model-based, nor validated on a realistic environment of peptide-centric protein inference.

Problems unresolved in protein inference workflow are, number of proteins to consider for reference, quantifying degenerate/shared peptides, dealing with ambiguous protein identifications and reducing the error rate at the protein level (since the error rate at protein level is substantially higher than that at the peptide level) [5] [7]. We base our inference algorithms on class properties that permit the occurrence of noisy sequence data: strength of class membership is defined as the absolute probability of identifying a member protein for a given reference protein. Our objective is to accelerate the computational workflow of protein inference. The inference process has functions which execute offline, online and run only at the experimental setup. We focus only on the online functions on execution.

The rest of the paper is organized as follows: Section II, defines the problems addressed. Two previous attempts to develop the computational workflow of protein inference are presented in Section III. Section IV summarises our contribution and Section V explains the hardware-software co-design concepts used. We present our methodology of solving the computational problem in the inference workflow in Section VI. Section VII describes the experimental setup we developed. We compare our hardware accelerated results of inference workflow with respect to a software only system with 50MHz frequency in Section VIII. Finally, we conclude in Section IX.

## II. PROBLEM DEFINITION

In shotgun proteomics, inference refers to the process of finding the origin of each peptide sequence and finding which proteins are present in the sample. When a disease spreads a large number of protein samples of suspected disease affected biological bodies are collected. Digestion and mass spectrometric analysis of these protein samples can provide the peptide set that becomes the input to the inference process. Mapping these peptides to parent proteins and assigning confidence levels for identified proteins is a challenging task because of the high peptide generation rate and the ever-increasing size of protein databases against which this mapping has to be carried out. The existing software based solutions are not adequate to handle this problem efficiently.

In general, a given protein cannot be defined by a fixed sequence of amino acids due to the existence of splice variants and isoforms of that protein. There are two problems that arise from these mutations: a) due to the permitted variations, the applicability of exact string matching methodologies could be questioned, b) the difficulty of defining a reference sequence for a particular set of proteins that are functionally indistinguishable, but with some variation in features. Fig. 1 shows an example of the alignment made with ClustalW tool [27] of three protein sequences having some common functional properties (but with differing lengths) which is permitted in some positions (indicated in the alignment in the last row using the symbols '.', ':' and '*', which stand for, conservation between groups of weakly similar properties, conservation between groups of strongly similar properties and a fully conserved residue, respectively). Our methodology to address these latter issues will be discussed in the sequel paper.

Another variant of the inference problem arises in the case of a complex protein sample without adequate evidence to focus the search. Here the search should extend into the complete database of proteins, making the process computationally very expensive. The problem becomes more pronounced when the search has to be done online.

### III. RELATED WORK

Much research has been done on faster solutions to the protein inference problem, mostly by accelerating the computational workflow of protein. PeptideClassifier [5] is a software based implementation which extracts the identified peptide sequences by a search engine (such as Mascot/SEQUEST/X!Tandem) and classifies them into six (eukaryotes) and three (prokaryotes) predefined evidence classes. Its major concern is to handle the peptide degeneracy problem and minimizing the propagation rate of the error by the gene model–protein sequence–protein identifier based classification that is introduced [5]. De-novo protein identification approach was introduced by Alex et al. [6] in order to identify a completely new protein by looking at the genomes already identified. Its major concern is to make the genome search faster by performing a parallel search with multiple copies of a particular genome. They match a single nucleotide sequence that is derived from a peptide at a time, but in parallel and exhaustively (for all possible combinations of nucleotide sequences). Further they have implemented the complete set of processes in the inference workflow in hardware [6]. They have achieved 30 times speed-up only for the matching process at gene level compared to an implementation running on a modern PC. An un-optimised co-designed architecture was used in our previous work [28], and in [29] we introduced co-design method for protein identification applications that measured the scalability of the system with the increase of the number of patterns. Tandem mass spectrometry based identifications of protein variations have been reported in the past [22][23]. Several researchers have worked on approximate or tolerance allowed string matching, but not on Aho-Corasick (A-C) with limited tolerance, to our knowledge. Alexei Nevidomski et al. introduced a method for approximate matching with the use of trie data structure in 2006 [24].

### IV. OUR CONTRIBUTION

The major contribution of our work is the optimised model-based hardware-software co-design methodology developed and validated for protein inference. This method partially eliminates the need for searching through a huge database and reduces the search time. We have first developed our methodology based on reference proteins in clustered databases. Then this has been extended to a hardware accelerated methodology for an inference workflow that accepts some tolerance in inference process for incoming proteins in complex mixtures of large volumes. This is applicable for a complex protein sample without a-priori knowledge (evidence) that could help in limiting the scope of search and for high throughput mass-spectrum identifications of peptides from patients of known/unknown diseases.

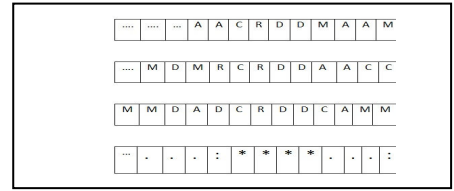

Fig. 1. Alignment to Fig. out a reference protein for a protein cluster

Identification of Baylisascaris Procyonis (B.P.) infection was used as an application instance. This workflow can be generalised to any inexact multiple pattern matching application by replacing the patterns in a clustered and distributed environment which permits a distance between member strings to account for permitted deviations such as substitutions, insertions and deletions. A methodology is also proposed to infer proteins by taking several UniRef database clusters at a time to run on an FPGA in parallel logic units. To address the possibility of bottleneck formation while performing alignments with ClustalW tool (ClustalW-MPI [26]), due to the resulting larger set of proteins, the experimental setup is extended by increasing the number of nodes that runs ClustalW in a distributed environment with the use of a set of FPGAs. A detailed description of the proposed model-based hardware-software co-design approach is provided, from specification to implementation and validation. Significant performance speed-ups have been achieved compared to the software-only approach and this is expected to improve still further with the new limited-tolerance based approach being tested at this moment.

### V. HARDWARE-SOFTWARE CO-DESIGN

The current technology favours hardware- software co-designed implementations for performance-critical workflow/algorithms [9]: either the design is first implemented in software and profiling steps move the critical parts to hardware or it is made for hardware and possible parts are moved to software till the design constraints are met. Major activities involved in the process of hardware-software co-design are: Specification and modelling, Design and Validation [10] (Fig. 2) [11]. Specification can be of several types: behavioural, architectural, requirements and executable specifications [12]. Behavioural specification relates to the functional parts of the system including algorithms and control flow [12]. Architectural specification defines the architectural parts of the systems that are reused such as processors and IP components [12]. Requirements specification addresses non-functional system requirements such as throughput constraints, power consumption constraints and descriptions of external interfaces [12]. Executable specification relates to the way the system behaviour is converted into some executable format in a computer, so that verification and validation of the system could be performed. Language based specification could use either homogeneous modelling or heterogeneous modelling. Homogeneous modelling uses a single programming language for developing both hardware and software (e.g. VHDL or C), whereas heterogeneous approach uses different languages for hardware (e.g. VHDL) and software (e.g. C) [14]. In this latter case, mapping to hardware and software is simple, but interfacing and validation is difficult. Design phase of the co-design consists of several steps: task assignment, cost estimation, hardware-software partitioning, co-synthesis, integration and implementation. Task assignment should be performed after properly completing the specification and involves dividing the system specification according to the basic tasks/ functional modules that satisfy the system functionality. The cost estimation could be done in order to measure power, time, communication, etc. We are mainly concerned with time. Co-synthesis includes communication synthesis, specification refinement, hardware synthesis and software synthesis. The process ends with system validation.

## VI. METHODOLOGY

Our approach starts from an examination of the known set of splice variants and isoforms of a target protein and the reference protein for a given protein cluster. For the situation where inference has to be carried out on a complex protein sample without any a-priori clue about target reference proteins, we propose a methodology for extracting several clusters at a time in a distributed environment where each processing node runs an instance of ClustalW-MPI program. This parallel processing should discover the reference protein faster. Next, this reference protein could be digested to obtain its corresponding peptides for generating A-C automata. The automata generated by different nodes should be uploaded to FPGAs without any delay in order to perform the matching with the unknown input protein mixture which is already digested and identified. Fig. 4 shows the designed environment for running inference process in a distributed environment. This method of defining a reference protein from defined clusters in UniRef is meaningful when considering common properties of proteins in each cluster.

Our inference system was designed to infer proteins using an estimate of the absolute probabilities of identifying all constituent peptides of the respective proteins (probability of identification in the presence of isoforms and splice variants may not be 100%) in digestion products of an input protein sample. In this context, the behavioural specification includes functions such as controlling input (extracting sequences from FASTA formatted files and making random protein samples) and output (identified proteins with probabilities), protein digestion algorithm, selecting highly significant peptides (if the output of Mascot program is used as the input to our system), A-C algorithm for peptide matching, encoding peptides/proteins, mapping peptides to proteins, calculation of absolute probability for protein identification and calculation of false/true discovery rates. The requirement specification covers accelerating the protein inference process: time is the most important constraint here. To suit the storage capacity of our FPGA we reduced our scope to identify a specific infection (eg: B.P. infection) to constrain the number of logic elements it consumed for the A-C logic. We have used the Avalon Memory Mapped interface provided by Altera in order to connect peripherals on a system on a programmable chip (SOPC) environment [13]. The choice regarding the executable specification was a heterogeneous system and time was the only metric to accelerate the inference workflow. The tasks identified in our inference system assigned to the modules during task assignment phase were: input output management, protein digestion, peptide selection, etc. Input output management includes functions such as extracting data from the FASTA files offline to make protein samples as the input to the digestion algorithm, writing the inferred proteins to the console of the Nios II IDE with probabilities and writing the false discovery rate/true discovery rate to the console of the Nios II IDE. In performing in-silico digestion of mitochondrion proteins of B.P. we followed the rules [15]: trypsin recognizes the basic amino acids lysine (K) and arginine (R) and cleaves carboxy-terminally (K or R in position P1 in Fig. 3), cleavage is refused if there is proline (P) in position P1' and trypsin of higher specificity additionally does not cleave after K in CKY, DKD, CKH, CKD, KKR nor after R in RRH, RRR, CRK, DRD, RRF, KRR [15]. Fig. 5 shows an example of a set of peptides obtained from a protein segment after trypsin digestion. We used likelihood probabilities (probability of identifying a peptide incorrectly) of peptides for peptide selection; if this likelihood probability value for a peptide is larger than 0.1 [16] it is not considered as a significant peptide. Multiple pattern matching algorithms (e.g. A-C) were used to match multiple peptide patterns in a single pass. This identifies clustered peptides (proteins are extracted from mitochondrion of B.P. proteins in UniRef clusters and each peptide extracted is mapped to a particular protein in the UniRef clusters) into categorized automata (several automata are used here representing a maximum of 32 peptides in one automaton that is selected to the optimisation algorithm we developed earlier [21]). In the software implementation we have used Multifast library [17] and modified it for developing our pattern specific automata. In the hardware implementation we have used bit-split [18] version of A-C algorithm. Peptide and protein encoding is used in hardware implementation and array indexing in software implementation. For each peptide represented in A-C automata, there is a protein of origin which should be kept as a map (Peptide protein mapping) because while constructing automata we have shuffled all the peptides according to the peptide reordering algorithm. After identifying peptides an estimate of the absolute probability ($\pi$) can be obtained using (1) below: calculated with respect to the maximum number of peptides ($\beta$) that could be present in each protein. Here $\alpha$ refers to the number of peptides identified from the cluster. Selecting high confidence proteins is related to the real workflow of protein inference and it is performed after identifying the sequence of peptides present in the sample via Mascot tool.

$$\pi = \alpha/\beta. \quad (1)$$

$$\lambda = \sigma/\phi. \quad (2)$$

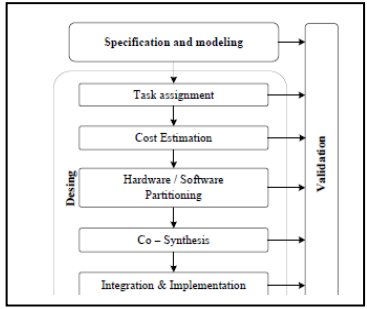

Fig. 2. Activities of Hardware/Software Co-design

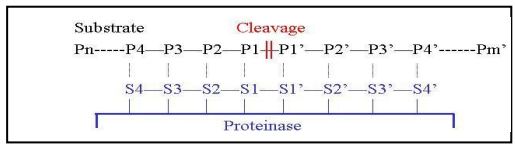

Fig. 3. Cleavage Specificity

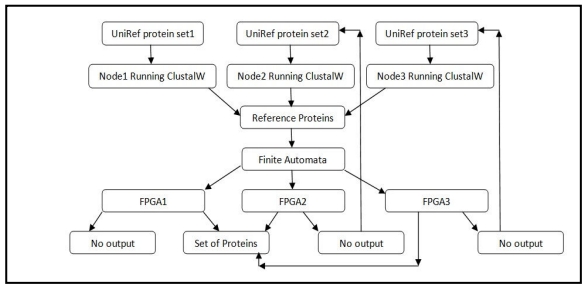

Fig. 4. Design of distributed environment for protein inference

Prior to performing co-design, we identified the time critical functions of the computational workflow of protein inference by performing profiling on Nios II processor, because profiling on a different system may affect the instruction set architecture and potential optimisations and finally result in variances in the executable of profiling. There exist other profilers that use either instrumentation or sampling methodology; AMD CodeAnalyst [19] and gprof. This is good for CPU bound problems. Valgrind performs profiling by instrumentation methodology [8] [20] and it suits the problems based on input, output and memory better. We performed profiling on Nios II processor with the use of performance counter for several different data sets to decide which portions of the workflow were to be moved to hardware. The ratios are presented here for the online functions in the real inference workflow that include setting the peptide input, peptide search, protein-peptide mapping and absolute probability calculation for the identification of each reference cluster. Then we moved hot-spots to hardware leaving other functions in software. The software only system was implemented to run on Intel Core i7 processor as well as Nios II processor for the functional modules identified in the behavioural specification and the tasks identified in task assignment phase.

Peptide search and protein-peptide mapping functions take 95.7%-97.2% time with respect to the time taken in entire workflow, other online functions including setting the peptide inputs and absolute probability calculation and input handling take 0.31%-1.37% and 1.9%-3.5% time, respectively, with respect to the total time taken. Fig. 6 gives a summary of profiling results on Nios II processor.

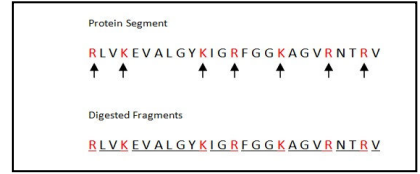

Fig. 5. Peptides after trypsin digestion

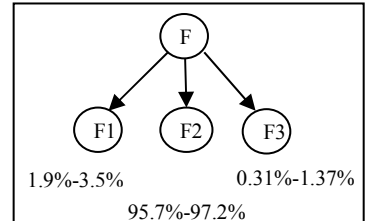

Fig. 6. Functional dependency graph :F-main, F1-input handling, F2- peptide matching and protein-peptide mapping, F3- probability calculation

According to the profiling results, we have moved 95.7%-97.2% of functions to the hardware and the remaining part was kept in software (run on Nios II processor). The maximum theoretical speedup that the whole hardware software system could obtain according to the Amdahl's law was computed for these ratios. The theoretical speed-up of co-designed system was measured with respect to a similar software only system running on Nios II processor (time taken in software only system/time taken only for software portion in co-designed system). According to these the entire inference flow should have 23.6 times to 35.7 times speed-up if we assume the time taken for hardware processes (matching and mapping) is 0μs. However hardware processes takes some time and it also includes the overhead of calling functions to measure time which is more than the time taken for processing one character/amino acid of a peptide.

Online and offline functions in the real workflow and experimental workflow are as given earlier. In the first implementation A-C machines were built for exact identification of peptides and identifying the protein cluster with some probability. In the new method that identify peptide sequences with limited tolerance (that may include functional properties) the number of finite state machines that need to be constructed gets significantly reduced since the regions in peptides which are not conserved are ignored. The basic assumption here is that the variations always happen inside peptides. For example, for the alignments made in the Fig. 2, a segment of that protein cluster can be represented by ($)(MD)(A|M)(A|R|D)CRDD(MAC)(A|D)(M|A|C)(M|C).
Here (A|M) means that residue only has either 'A' or 'M' amino acids. If we were to make finite state machines for all the combinations in Fig. 2 we have to make many more A-C

machines (in fact machines to identify 72 times more peptide sequences) and it will consume much more storage in FPGA. Therefore, by this method we could have a higher speed-up than in our previous method due to FSM reduction. There also exist functional units and domains of proteins for some known proteins in Prosite database [25], which will also be included, whenever possible, to make an additional search to confirm whether the results are consistent with these domains after the first search is done.

## VII. EXPERIMENTAL SET UP

B.P. Infection was used as the disease instance for performing inference experiments. Diagnosis of this infection through morphologic identification is now obsolete. Currently, there exist several methods of diagnosis including (a) serologic tests to examine the antigen developments by specific antibodies ES and BpRAG1, (b) use of clinical findings or exposure history, (c) urine test or stool test or blood test to find the exact parasite causing the infection (basically whether it is a flatworm or roundworm). The flatworms include flukes, tapeworms and the roundworms include pinworms, threadworms and hookworms. Our method could assist in performing (c) on a protein sample from urine/stool/ blood of host/ infected organism .The samples can be created as mentioned in [31] for this kind of analysis. This will help diagnosis and control of the disease by identifying the parasite correctly from the mitochondrion proteins of B.P. This could be further used to design drugs in pharmaceutical industry by looking at the presence of peptides in the sample. Further analysis with protein biomarkers of the B.P. might help to differentiate between other Baylisascaris species which are evolutionary related.

The samples were prepared artificially: FASTA formatted clustered mitochondrion protein data of Raccoon Roundworm (member sequences from all clusters) were obtained from UniRef database where clusters have similar proteins allowing some distance between the reference protein and each member protein. In this database there are 12 clusters corresponding to 12 protein coding genes of Raccoon Roundworm. Sequence data were extracted from a FASTA formatted file and written to a tab delimited text file along with protein cluster id. Next, proteins were randomly selected from different clusters from this protein member data and digested using our in-silico protein digestion tool (an off-line adaptation of PeptideMass [15] that was developed for the purpose). From this the artificial complex peptide sample to be identified was prepared. In the real workflow there would be an identified peptide sample (from Mascot tool) instead. In the present application the concentration of each peptide present in the sample is not measured in order to find number of proteins from each cluster. Therefore only one protein was used from each cluster, hence, resulting in a maximum of 12 clusters. While preparing the samples artificially we do not consider the number of members in each cluster; proteins are selected randomly. The incidence of both over-sampling and under-sampling can be avoided by considering properties: the concentration of each cluster as well as peptide detectability of a peptide while making samples.

In order to make the computational workflow faster, a minimum list of known reference proteins which do not contain degenerate peptides that helps identification of B.P. infection was used. However, members of a particular cluster referred by a reference protein could exhibit some degenerate peptides of an isoform or splice variant of a protein /other homologous protein. Since we have clusters with 0.5 and 0.9 identities (similarity between two protein sequences: reference sequence and member sequence), we could consider that members of the cluster have evolutionary change (substitution, deletion and additions of amino acids) with respect to the reference sequence. This means "identity number" of each cluster represents the amount of deviation of each sequence from reference protein sequence.

FSM initialization and creation was done offline from the digested reference proteins in UniRef clusters. For this in-silico digestion was done with the developed tool allowing mentioned digestion rules. Then the peptides were arranged into a new order and a new categorisation made according to our optimisation algorithm presented in [21]. Later, protein-peptide mapping was performed offline. Next, the FSMs were generated for the peptides obtained. Input setting, peptide search and identification, probability calculation of identified peptides for each cluster, validation with false discovery rate are done online where validation process is included only in the experimental workflow. The software only implementation and the software portion of co-designed implementation were run on the Nios II processor with 50MHz frequency. Hardware implementation was run on Altera Cyclone II FPGA.

TABLE I. COMPARISON OF RESULTS

| No. of proteins in the sample | Average Time(μs) of 30 samples | | | |
|---|---|---|---|---|
| | *SW Only Implementation on Nios II processor* | *Subhead HW SW co-designed implementation* | *Average speed-up* | *Expected theoretical speed-up* |
| 2 | 7959.06 | 1300.83 | 6.12 | 23.64 |
| 4 | 16557.3 | 2656.7 | 6.23 | 25.48 |
| 6 | 25863.23 | 4015.5 | 6.44 | 24.21 |
| 8 | 32528.53 | 11797.53 | 2.76 | 28.08 |
| 10 | 54390.93 | 9462.8 | 5.75 | 35.71 |
| 12 | 86538.3 | 8713.67 | 9.93 | 30.75 |

## VIII. RESULTS

Average times taken for on-line computations by different, randomly selected samples of 30 proteins are shown in Table I for both implementations (software implementation and HW/SW co-designed implementation). It also shows the expected theoretical speed-ups and the observed average speed-up for 30 samples. Maximum speed-up observed was 10 (9.9) and minimum was 3 (2.7). Here we could not get expected speed-up and linearity due to several reasons: a) difficulty in reducing timer overhead taken by timestamp routines when measuring time on Nios II, b) when the target slave is busy reading the input and writing the output to the register, number of clock cycles taken by Memory Mapped Slave may vary, c) running code from SDRAM through instruction cache will give different timing depending on whether the code is cache resident or not and d) in the worst case the program may access the host file system to keep peptide data and generated automata when the cache is not enough. Here a) and b) contribute more to the limitation of the speed-up because we don't have any other way of profiling a Nios-II system except using the time-stamp routines and we read from and write to the same register (register1) in Avalon Memory Mapped interface. Reason c) also seems to contribute here, because sometimes we notice different timings for the same set of protein inputs. Reason d) could contribute to the results rarely. Here we don't have such instances.

We observe that the correct inference of proteins ranges from 50% to 100% in both co-designed and software only implementations. It is due to the distance between reference and member proteins in clusters. For a given suspected protein sample, currently we identify the relevant protein clusters that represent 12 proteins of protein coding genes expressed in mitochondrion of roundworm. As future work, a model could be developed to identify the correct member from each cluster. Currently we do it by looking at the probability of protein identification. There could be situations having exactly the same probability for several members in the cluster. Models should be introduced for this purpose. Finally, the runs we are currently conducting on our new algorithm for A-C machines with limited tolerance (reported in a sequel paper) should provide much elevated speed-ups and we intend to make a performance comparison between these tests and the ones given above in a subsequent publication.

IX. CONCLUSION

The model based inference system presented permits protein-peptide mapping without searching through a huge database. This helps saving time in assembly process. Profiling helps identifying hot-spots in the software based system. Currently we are developing a generalized system for identifying functionally conserved regions for a disease by reducing the solution space with the concept of greatest common set of sub-strings allowing limited tolerance in the peptide sequences at peptides level. This approach is described in the sequel paper.

REFERENCES

[1] Y. F. Li and P. Radivojac,"Computational approaches to protein inference in shotgun proteomics", BMC Bioinformatics 2012, vol. 13, 2012.

[2] A. I. Nesvizhskii and R. Aebersold, "Interpretation of shotgun proteomic data: the protein inference problem", Molecular and Cellular Proteomics, 2005, pp. 1419-1440.

[3] J. Shi, "Protein inference based on peptides identified from tandem mass spectra", PhD Thesis, University of Saskatchewan, 2012.

[4] L. Otvos, "Peptide-Based Drug Design: Here and Now", Peptide-Based Drug Design Methods in Molecular Biology,vol. 494, 2008, pp. 1-8.

[5] E. Qeli and C. H. Ahrens,"PeptideClassifier for protein inference and targeted quantitative proteomics", Nature Biotechnology, 2010, pp. 647-650.

[6] A. Alex, J. Rose,R. Isserlin-Weinberger and C. Hogue, "Hardware Accelerated Novel Protein Identification", Field Programmable Logic and Application- Lecture Notes in Computer Science, vol. 3203, 2004, pp. 13-22.

[7] K. M. Arendt , W. M. Old , S. Houel, K. Renganathan, B. Eichelberger, K. A. Resing, and N. G. Ahn, "IsoformResolver: A Peptide-Centric Algorithm for Protein Inference", Journal of Proteome Research, 2011, pp.3060-3075.

[8] *Understanding Profiling Methods*, [Online]. Available: http://msdn.microsoft.com/en-us/library/dd264994.aspx, Microsoft Developer Network, 2014.

[9] R. Joost and R. Salomon, "Hardware-Software Co-Design in Practice:A Case Study in Image Processing", 32nd Annual Conference on IEEE Industrial Electronics, 2006, pp.3674-3679

[10] A. Shaout, A. H. El-Mousa., and K. Mattar, "Specification and Modelling of HW/SW Co-Design for Heterogeneous Embedded Systems," Proceedings of the World Congress on Engineering, vol. 1, 2009.

[11] N. Bencheva, N. Kostadinov , Y. Ruseva, "On Teaching Hardware/Software Co-design using FPGA," Lecture notes, Ruse University, 2010.

[12] M. O'Nils, "Specification, Synthesis and Validation of Hardware/Software Interfaces," Doctoral Thesis, Electronic system Design, Department of Electronics, 1999.

[13] Altera Corporation, "Avalon Memory-Mapped Interface Specification," 2007.

[14] B. S. El-Haik, A. Shaout, "Software Design for Six Sigma: A Roadmap for Excellence," 2010.

[15] *PeptideMass*, [Online]. Available: ttp://web.expasy.org/peptide_mass/ peptide-mass-doc.html#table1, 2014.

[16] *Scoring*,[Online].Available:http://www.matrixscience.com/help/scoring_ help.html, Matrix Science, 2014.

[17] *Multifast*, [Online]. Available: http://multifast.sourceforge.net/, 2013.

[18] Y. S. Dandass, S. C. Burgess, M. Lawrence and S. M. Bridges, "Accelerating String Set Matching in FPGA Hardware for Bioinformatics Research", BMC Bioinformatics,vol. 9, 2008.

[19] *CodeAnalyst*, [Online]. Available: http://developer.amd.com/tools-and-sdks/archive/amd-codeanalyst-performance-analyzer/, 2014.

[20] *Sampling Vs Instrumentation*, [Online]. Available: https://code.google.com/p/oktechprofiler/wiki/SamplingVsInstrumentati on, 2014.

[21] S. M. Vidanagamachchi, S. D. Dewasurendra, R. G. Ragel and M. Niranjan, "Tile optimisation for area in FPGA based hardware acceleration of peptide identification", 6th IEEE International Conference on Industrial and Information Systems, 2011, pp. 140-145.

[22] C. L. Gatlin ,J. K. Eng , S. T. Cross , J. C. Detter , and J. R. Yates, "Automated identification of amino acid sequence variations in proteins by HPLC/microspray tandem mass spectrometry",Analytical Chemistry, 2000, pp. 757–763.

[23] P. A. Pevzner, V. Dančík, and C. L. Tang, "Mutation-tolerant protein identification by mass spectrometry",Journal of Computational Biology, 2000, pp. 777-787.

[24] A. Nevidomski and P. Volkov, "Method and system for approximate string matching", [Online]. Available: http://www.google.com/patents/ US20060004744, 2006.

[25] *Prosite*, [Online]. Available: http://prosite.expasy.org/, 2014.

[26] K.B. Li, "ClustalW-MPI: ClustalW analysis using distributed and parallel computing", Bioinformatics, vol. 19, 2003, pp. 1585-1586

[27] *ClustalW*, [Online]. Available: http://www.ebi.ac.uk/Tools/msa/ clustalw2/, 2014.

[28] S. M. Vidanagamachchi, S. D. Dewasurendra and R. G. Ragel, "Hardware Accelerated Protein Inference Framework", 8th IEEE International Conference on Industrial and Information Systems, 2013, pp. 649 – 653.

[29] S. M. Vidanagamachchi, S. D. Dewasurendra and R. G. Ragel, "Hardware software co-design of the Aho-Corasick algorithm: scalable for protein identification?", 8th IEEE International Conference on Industrial and Information Systems, 2013, pp. 321 – 325.

[30] E. L. Blizzard, C. D. Davis, S. Henke, D. B. Long, C. A. Hall and M. J. Yabsley, "Distribution, Prevalence, and Genetic Characterization of Baylisascaris procyonis in Selected Areas of Georgia",Journal of Parasitology,2010, pp. 1128-1133.